\documentclass[a4paper,fleqn]{cas-dc}
\usepackage[numbers]{natbib}
\usepackage{subfigure}
\usepackage{epstopdf}

\def\tsc#1{\csdef{#1}{\textsc{\lowercase{#1}}\xspace}}
\tsc{WGM}
\tsc{QE}
\tsc{EP}
\tsc{PMS}
\tsc{BEC}
\tsc{DE}

\begin{document}
\let\WriteBookmarks\relax
\def\floatpagepagefraction{1}
\def\textpagefraction{.001}
\shorttitle{Learned Image Compression with Generalized Octave Convolution and Cross-Resolution Parameter Estimation}
\shortauthors{H. Fu et~al.}

\title [mode = title]{Learned Image Compression with Generalized Octave Convolution and Cross-Resolution Parameter Estimation}                      
\tnotemark[1]

\tnotetext[1]{This work was supported in part by the National Natural Science Foundation of China (No. 61474093), in part by the Natural Science Foundation of Shaanxi Province, China (No. 2020JM-006), and in part by funding from the China Scholarship Council (CSC).}

\author[1]{Haisheng Fu}
\author[1]{ Feng Liang}[orcid=0000-0002-9393-6224]]
\cormark[1]
\ead{fengliang@xjtu.edu.cn}

\address[1]{School of Microelectronics, Xi'an Jiaotong University,  China}

\begin{abstract}
Recently, image compression approaches based on deep learning have gradually outperformed existing image compression standards including BPG and  VVC intra coding. In particular, the application of the context-adaptive entropy model significantly improves the rate-distortion (R-D) performance, in which hyperpriors and autoregressive models are jointly utilized to effectively capture the spatial redundancy of the latent representations. However, the latent representations still contain some spatial correlations. In addition, these methods based on the context-adaptive entropy model cannot be accelerated in the decoding process by parallel computing devices, e.g. FPGA or GPU. To alleviate these limitations, we propose a learned multi-resolution image compression framework, which exploits the recently developed octave convolutions to factorize the latent representations into the high-resolution (HR) and low-resolution (LR) parts, similar to wavelet transform, which further improves the R-D performance. To speed up the decoding, our scheme does not use context-adaptive entropy model. Instead, we exploit an additional hyper layer including hyper encoder and hyper decoder to further remove the spatial redundancy of the latent representation. Moreover, the cross-resolution parameter estimation (CRPE) is introduced into the proposed framework to enhance the flow of information and further improve the rate-distortion performance.  An additional information-fidelity loss is proposed to the total loss function to adjust the contribution of the LR part to the final bit stream. Experimental results show that our method separately reduces  the decoding time by approximately 73.35 $\%$ and 93.44 $\%$ compared with that of state-of-the-art learned image compression
methods, and the R-D performance is still better than H.266/VVC(4:2:0) and some learning-based methods on both PSNR and MS-SSIM metrics across a wide bit rates.
\end{abstract}

\begin{graphicalabstract}
\includegraphics[scale=.75]{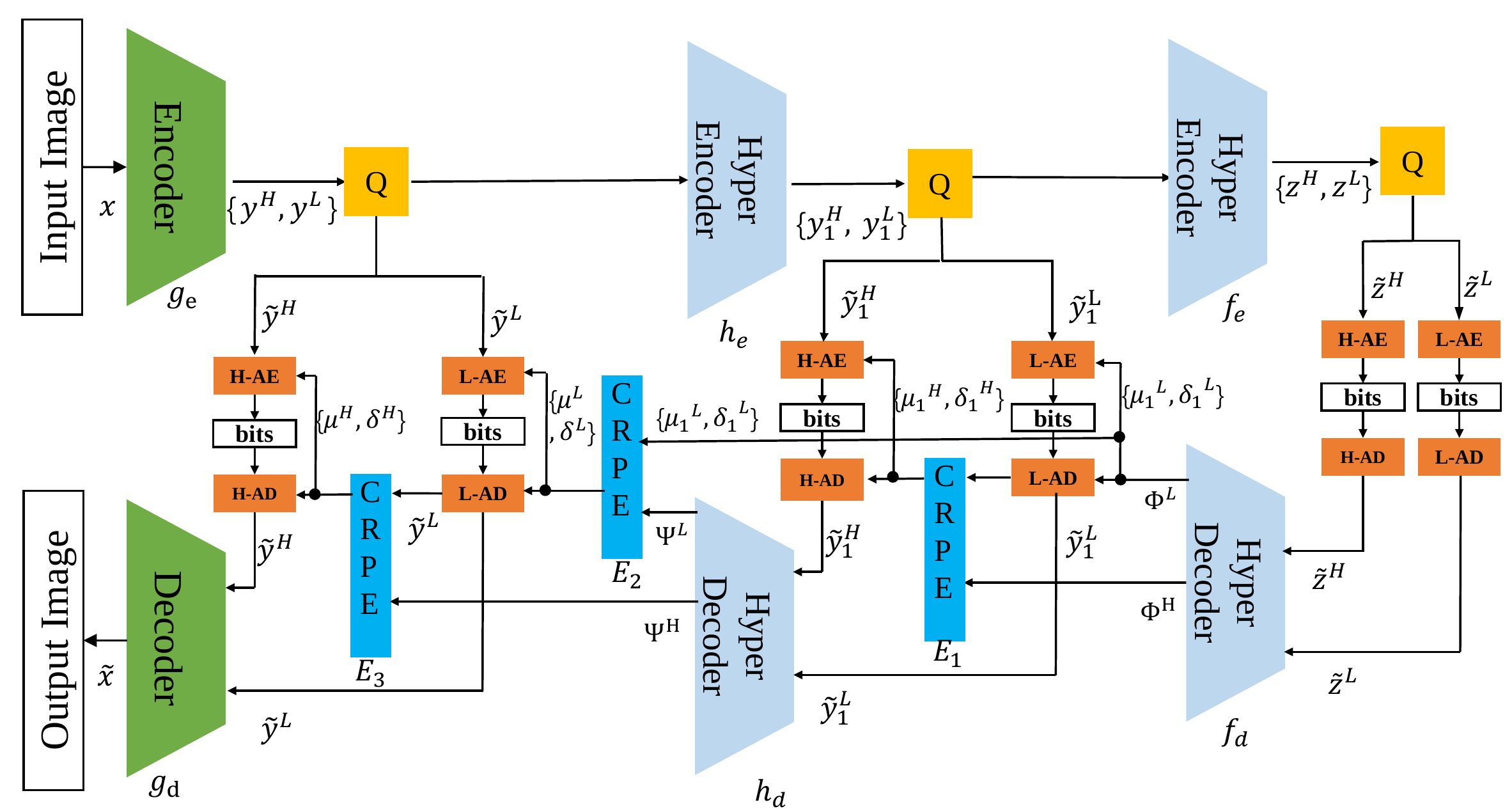}
\end{graphicalabstract}

\begin{highlights}
\item Learned Image Compression
\item Generalized Octave Convolution 
\item Outperforms the state-of-the-art deep learning-based methods and traditional codecs such as VVC
\end{highlights}



\maketitle

\section{Introduction}
Image compression is a crucial and fundamental topic in the field of multimedia communications. Traditional image compression standards including JPEG \cite{JPEG}, JPEG 2000 \cite{JEPE2000_book}, and BPG (intra-frame encoding of HEVC/H.265) \cite{BPG} are mainly composed of the following components$:$ linear transform, quantization, and entropy coding.

Recently, image compression approaches based on deep neural networks (DNNs) have achieved competitive results and have gradually outperformed existing image compression standards including BPG and the upcoming VVC intra coding. The end-to-end learned image compression approaches aim to minimize the rate-distortion (R-D) cost. However, it is difficult to establish accurate and differentiable estimates of the entropy. One of the most common schemes is to assume that the symbol streams are independent and identically distributed (i.i.d.), and then use parametric models to estimate the distribution of the symbols. In \cite{Lossy_image, soft_to_hard, Real_time}, different end-to-end learned image compression frameworks based on DNNs were presented,  and all the symbols are assumed to be i.i.d..  Since the entropy models are not established accurately enough, the performances of these methods are limited. To establish a more accurate entropy model, some image compression approaches based on the hyper-prior model and joint model were proposed.  In \cite{Joint}, the autoregressive information was extracted by 2D $5\times5$ shielding convolution for probability estimation, which is the first approach to outperform  BPG (4:4:4) in PSNR. In \cite{cheng2020},  Cheng et al. proposed the discretized Gaussian Mixture Likelihoods (GMMs) module to parameterize the distributions of the latent representations, which can exploit a more efficient and flexible entropy model to capture spatial dependencies of the latent representations.


Many learned image compression methods exploit information enhancement methods to further improve the R-D performance. In \cite{GDNGDN}, the generalized divisive normalization (GDN) operator was introduced  into the transformation function for joint nonlinearity and spatial adaptability, which was proven to be very suitable for natural image compression.  In \cite{cheng2020}, the simplified attention module is introduced to improve the R-D performance, which is easier to obtain global information and reduce time complexity. In \cite{channel}, two enhanced modules including channel-conditioning module and latent residual prediction module are presented, which lead to network architectures with better R-D performance than some existing context-adaptive learned schemes while minimizing serial processing.

Different approaches have been proposed to design an efficient and effective network structure to improve the R-D performance of image compression. The designed structure can extract more accurate and flexible latent representations of the input image and obtain high-quality decompressed images from the quantized compressed feature. For example, in \cite{Pixel_RNN, RNN_2015, RNN_2017}, the recurrent neural networks were presented to compress the residual information recursively, which mainly relied on binary representation at each iteration to achieve scalable coding. In \cite{Real_time, GAN_2018, GAN_Generative}, the generative models were exploited to learn the distribution of input images, which can achieve better subjective quality at extremely low bit rates by adversarial training.

Most DNN-based image compression schemes \cite{soft_to_hard, GDNGDN, FU_2020, FU2021, Conditional, GDN, Pixel_RNN, Lossy_image, Real_time} are based on the autoencoder framework to extract a low-dimensional latent representation of the input image. In \cite{Variational}, Ball\'{e} et al. proposed a two-layer approach, which introduces the hyper layer on top of the autoencoder layer. The hyper layer (including a hyper encoder and a hyper decoder) can capture spatial dependencies among the elements of the latent representation of the autoencoder layer. Based on \cite{Variational}, in \cite{Hu_AAAI}, a coarse-to-fine framework with multi-layers was proposed to  effectively remove spatial redundancy, which achieved better performance than \cite{Variational}. Also, the information aggregation reconstruction sub-network is designed for reconstruction quality, which combines the output of the main decoder and the two hyper decoders to generate the final reconstructed image. The information aggregation module will significantly increase network complexity and computational complexity. In this paper, we just adopt the  multi-layer hyper-prior method to our framework. And other modules are introduced to take the place of the information aggregation reconstruction sub-network in \cite{Hu_AAAI}. Based on \cite{Hu_AAAI}, our new framework can achieve better image compression performance and reduces the encoding and decoding time.


Although the approaches based on DNNs have achieved  promising results, there are still some rooms for improvement. First, all latent representations have the same resolution, but we know from wavelet-based image compression that multi-resolution schemes can have better R-D performance for natural images \cite{JEPE2000_book}. Second, in these context-adaptive entropy model approaches, the encoding of an coefficient depends on the previously decoded neighbor representations. This sequential processing cannot be accelerated by parallel computing devices, e.g. FPGA or GPU. The decoding processing will takes a particularly long time.

To address these two question, in this paper, we propose an effective and efficient framework, which can improve both the parallel decoding and R-D performance of image compression. We can achieve a good trade-off between rate-distortion performance and decoding time. The contributions of the proposed method can be summarized as follows:

$\bullet\,$ We combine the multi-layer hyper-prior method \cite{Hu_AAAI} and  the GoConv modules \cite{octave} to form a new network architecture, which can further reduce the spatial correlations among the latent representation and improve the R-D performance of image compression.

$\bullet\,$More advanced cross-resolution parameter estimation (CRPE) modules are introduced in our scheme to improve the performance, which contains two types. The first type is incorporated in between high and low resolutions, which mainly improves the flow of information. The second type of the CRPE module is in different layers, which can combine the low-resolution (LR) parts of different layers. It is easier to obtain the global information for better encoding and decoding. The signal flow of most image compression frameworks is unidirectional. These two CRPE modules can improve the flow of information, in which the decoded information is provided as side information to the to-be-decoded elements.

$\bullet\,$To adjust the proportion of the LR part in the output bit stream, an additional information-fidelity loss is  proposed and added to the total loss function. Without this information-fidelity loss, the LR part will only occupy a very small proportion of the whole bit stream, which means the LR part is not fully utilized.

$\bullet\,$Our method separately reduces the decoding time by approximately 73.35 $\%$ and 93.44 $\%$ compared with that of state-of-the-art learned image compression methods on Kodak dataset.

Since we do not use element-wise context model, the proposed method can be decoded in parallel. Thanks to the contributions above, our scheme still achieves comparable performance in both PSNR and MS-SSIM quality metrics across a wide range of bit rates, in comparison with the traditional image codecs including VVC (4:2:0) and some recent learned image compression methods on both Kodak dataset and Tecnick dataset.

The remainder of the paper is organized as follows. In Section  \ref{Related work},  We briefly overview some relevant methods. In Section  \ref{The Proposed structure}, we propose our multi-layer hyper-prior and cross-resolution parameter estimation image compression method. In Sec. \ref{Experiment}, we compare the proposed scheme with some  state-of-the-art learned compression methods and traditional methods. We also conduct a series of ablation experiments to verify the performance of our proposed scheme. The conclusions are given in Sec. \ref{Conclusion}.

\begin{figure*}
	\centering
		\includegraphics[scale=0.75]{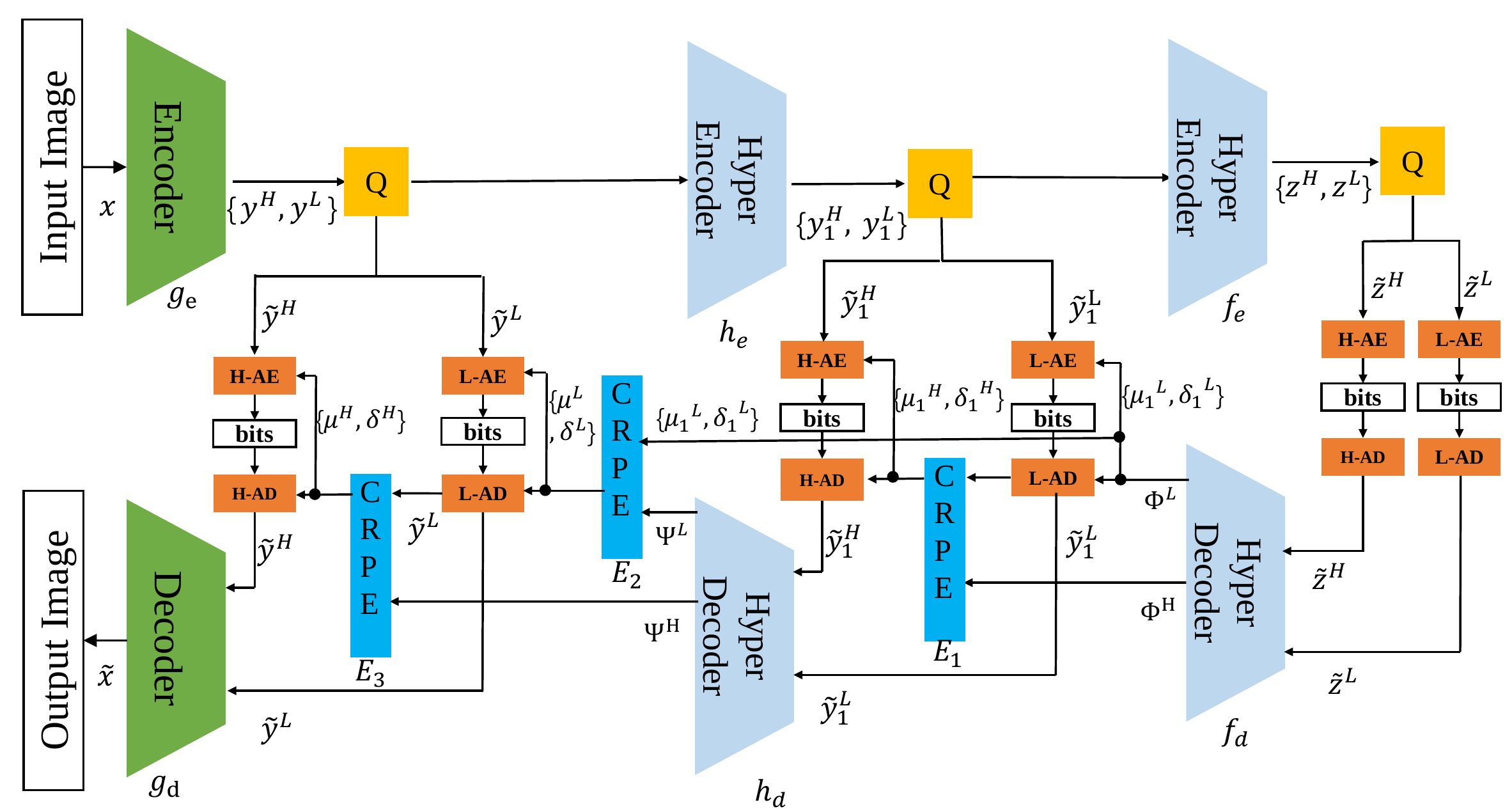}
	\caption{The proposed architecture with multi-layer hyper-priors and cross-resolution parameter estimation. $Q$ represents the quantification. $H$ and $L$ correspond to high resolution and low resolution, respectively. $AD$ and $AE$ correspond to Arithmetic Encoder and Arithmetic Decoder, respectively.  $CPRE$ denotes the proposed cross-resolution parameter estimation. Solid dots indicate that two signal lines are connected. }
	\label{scheme_structure}
\end{figure*}

\section{Related work}
\label{Related work}

\subsection{ Learned image compression}

Recently, the learned image compression methods mostly are based on  autoencoders which have achieved promising results and gradually outperform classical image codecs including VVC.  In \cite{Variational}, a hyperprior network is first proposed to estimate the model parameters of the latent representation, which can further capture the spatial correlation. It achieves comparable performance with  BPG (4:4:4) codec in terms of  the PSNR metric. To further reduce the redundancy of the latent representation, the context model is proposed in learned image compression framework.  In \cite{Joint}, an autoregressive mask convolution is further employed in order to get context-adaptive entropy coding. It is the first method which outperforms BPG (4:4:4) in PSNR. In \cite{Lee_2020}, two types of contexts, bit-consuming contexts and bit-free contexts, are further developed. In \cite{cheng2020} and \cite{GLLMM}, the performance is further improved by introducing advanced entropy coding models, such as GMM and GLLMM. Since these methods \cite{Joint, Lee_2020, cheng2020, GLLMM} utilize the context model to improve the R-D performance, the decoding time of bit-streams will be  a lot.

Many learned image compression schemes aim to design an efficient and effective network architecture,  which is  easier to reduce spatial redundancy in the image signal and to generate the high-quality reconstructed images from compressed features. Some approaches incorporate some advanced modules in a convolutional neural network to enhance the image compression performance, such as generalized divisive normalization (GDN) \cite{GDN}, residual blocks \cite{resblock}, non-local attention module \cite{NonLocal_attension}, coarse-to-fine  structure \cite{Hu_AAAI}.  Other methods utilize recurrent neural networks (RNNs) \cite{Toderici15} or generative adversarial networks (GAN) \cite{Santurkar_2018} to improve the performance.

\subsection{Octave Convolution}

In the classical convolution, all input and output feature vectors have the same spatial resolution. There is some spatial redundancy in low resolution information, which is not efficient on both memory and computation cost. To alleviate this issue, the octave convolution was first proposed in \cite{octave}, which factorizes the feature vectors into the high-resolution (HR) and low-resolution (LR) components, similar to wavelet transform \cite{JEPE2000_book}.  The octave convolution has been successfully employed in the field of many computer vision tasks \cite{octave, xu_2021_Octave, Lin_2022_Octave, Mahammand_AAAI, Lin_MMSP}.

In \cite{Mahammand_AAAI}, Akbari et al. proposed the generalized octave convolution (GoConv) and applied it to learned image compression.  To improve the image compression performance, the context-adaptive entropy model is also used in \cite{Mahammand_AAAI}. Therefore although the performance of \cite{Mahammand_AAAI} is pretty good, the decoding complexity is too expensive owing to the entropy model.  In \cite{Lin_MMSP}, Lin et al. simplify the framework by removing the context model from the network, at the price of reduced performance. The Lagrangian parameter is introduced in the GoConv modules to develop a variable-rate scheme, in which the bitrate range can be controlled more precisely.   As in \cite{Variational}, both \cite{Lin_MMSP} and \cite{Mahammand_AAAI} are based on two-layer  autoencoder models. However, these two methods just introduce the GoConv to take the place of the classical convolution, and they don't consider the effect between high-resolution (HR) and low-resolution (LR) components.

\section{The Proposed Image Compression Framework}

\label{The Proposed structure}
In this section, we first describe the overall architecture of the proposed framework, and then explain the details of major components, including the encoder/decoder networks, generalized octave convolution, the cross-resolution parameter estimation (CRPE) modules, and loss functions.

\subsection{The Overall Architecture}

The proposed framework is described in Fig. \ref{scheme_structure}. It is mainly composed of the core autoencoder, two layers of hyper-prior subnetworks, and three cross-resolution parameter estimation (CRPE) modules. The core autoencoder includes Encoder ($g_{e}$) and Decoder  ($g_{d}$). The first layer of hyper-prior subnetwork consists hyper Encoder ($h_{e}$) and hyper decoder ($h_{d}$). The second layer of hyper-prior subnetwork includes hyper encoder ($f_{e}$) and hyper decoder ($f_{d}$). Three cross-resolution parameter estimation (CRPE) modules contain CRPE module ($E_{1}$), CRPE ($E_{2}$), and CRPE ($E_{3}$).  The Encoder ($g_{e}$) is used to learn a quantized latent representation of the input image, while the entropy subnetwork ($h_{e}$) and ($f_{e}$) are expected to learn a probabilistic model over the quantized latent representations of the previous layer, which is utilized for entropy coding or sent to an information module. The notations are summarized in the Table \ref{denotes}.

\begin{table}
\caption{The input and output signals of the detailed module}
\begin{center}
  \begin{tabular}{ccccc}
  \hline
  \textbf{Module}& \textbf{Input Signal}  & \textbf{Output Signal}\\
  \hline
  \textbf{Encoder ($f_{e}$)} & $x$  & $\{y^{H},y^{L}\}$  \\
  \textbf{Hyper Encoder ($h_{e}$)} & $\{\tilde{y}^{H},\tilde{y}^{L}\}$  & $\{\tilde{y}_{1}^{H}, \tilde{y}_{1}^{L}\}$  \\
  \textbf{Hyper Encoder ($f_{e}$)} & $\{\tilde{y}_{1}^{H},\tilde{y}_{1}^{L}\}$  & $\{z^{H},z^{L}\}$   \\
  \textbf{Hyper Decoder ($f_{d}$)} & $\{\tilde{z}^{H}, \tilde{z}^{L}\}$  & $\Phi_{H}, \Phi_{L}$ \\
  \textbf{Hyper Decoder ($h_{d}$)} & $\{\tilde{y}_{1}^{H}, \tilde{y}_{1}^{L}\}$  & $\Psi_{H}, \Psi_{L}$  \\
  \textbf{Decoder ($f_{d}$)} & $\{\tilde{y}^{H},y^{L}\}$  & $\tilde{x}$ \\
  \textbf{CRPE ($E_{1}$)} & $\tilde{y}_{1}^{L}$, $\Phi_{H}$,    & $\{\mu_{1}^{H},\mu_{1}^{L}\}$  \\
  \textbf{CRPE ($E_{2}$)} & $\{\mu_{1}^{H},\mu_{1}^{L}\}$ , $\Psi_{L}$,    & $\{\mu_{1}^{H},\mu_{1}^{L}\}$  \\
  \textbf{CRPE ($E_{3}$)} & $\Psi_{L}$, $\tilde{y}^{H}$   & $\{\mu^{H},\mu^{L}\}$   \\
  \hline
\end{tabular}
\label{denotes}
\end{center}
\end{table}

\begin{figure*}
	\centering
		\includegraphics[scale=0.5]{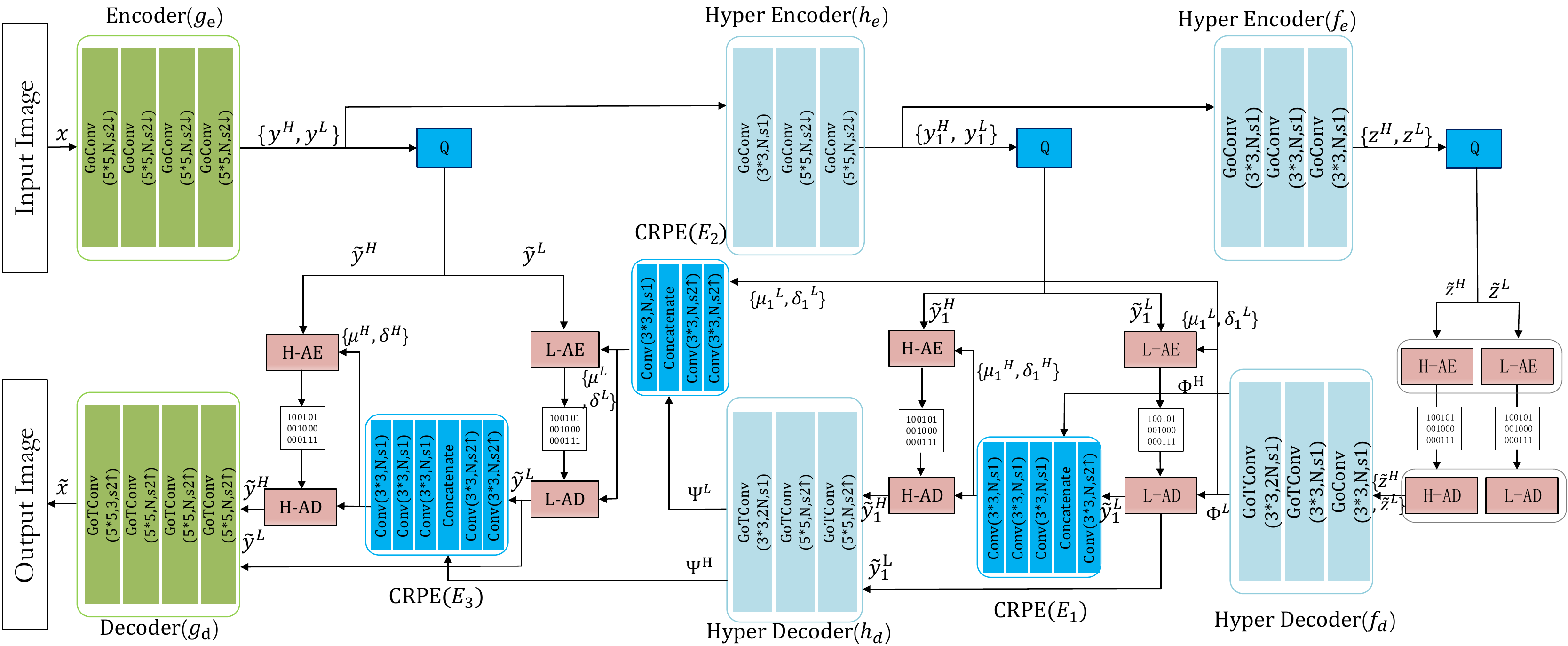}
	\caption{The detailed network architecture  of the proposed learned multi-resolution image compression based on generalized octave convolution network.}
	\label{framework}
\end{figure*}

In the encoding processing, the input image $x$ is transformed into the HR latent representation  $y^{H}$  and the LR latent representation $y^{L}$ by the Encoder network ($g_{e}$), which aims to reduce the pixel-wise redundancy of the input image. The details are explained in Sec. II.B. The GDN module in \cite{cheng2020} is adopted  as the activation in the Encoder network ($g_{e}$). The $y^{L}$ and $y^{H}$ be quantized as $\tilde{y}^{L}$ and $\tilde{y}^{H}$ respectively. To further reduce the redundancy within the latent representations, $\tilde{y}^{L}$ and $\tilde{y}^{H}$ are then further analyzed and sent to the hyper layers to further squeeze out the  redundancy. The $\tilde{y}_{1}^{L}$ and $\tilde{y}_{1}^{H}$ will be obtained by the hyper encoder ($h_{e}$). Last, $\tilde{y}_{1}^{L}$ and $\tilde{y}_{1}^{H}$ will be sent to the second hyper encoder ($f_{e}$) to obtain the $z^{L}$ and $z^{H}$, which will be quantized as $\tilde{z}^{L}$ and $\tilde{z}^{H}$ respectively. $\tilde{z}^{L}$ and $\tilde{z}^{H}$ will be encoded by factorized entropy encoder without building a prior probability model.  $\tilde{y}_{1}^{L}$, $\tilde{y}_{1}^{H}$, $\tilde{y}^{L}$ and $\tilde{y}^{H}$ are all coded by arithmetic coding. We estimate the probability distribution of $\tilde{y}_{1}^{L}$, $\tilde{y}_{1}^{H}$, $\tilde{y}^{L}$ and $\tilde{y}^{H}$ with the prediction model. The distribution of each element in $\tilde{y}_{1}^{L}$, $\tilde{y}_{1}^{H}$, $\tilde{y}^{L}$ and $\tilde{y}^{H}$ is assumed to be a Gaussian distribution with mean $\mu$ and variance $\sigma^2$.

In the decoding processing, as in \cite{Variational}, the quantized $\tilde{z}^{L}$ and $\tilde{z}^{H}$ will be obtained by the factorized entropy decoder without building a prior probability model. The quantized $\tilde{z}^{L}$ and $\tilde{z}^{H}$ will be sent to the second hyper decoder network to establish the distribution of each element in $\tilde{y}_{1}^{L}$. The decoded $\tilde{y}_{1}^{L}$ information will be added to the first CRPE module ($E_{1}$) to estimate the the distribution of each element in $\tilde{y}_{1}^{H}$. The decoded $\tilde{y}_{1}^{H}$ and $\tilde{y}_{1}^{L}$  will be sent to the first hyper decoder $h_{d}$ to obtain $\psi_{h}$. The $\psi_{h}$ and the decoded $\tilde{y}_{1}^{L}$ will be sent to the CRPE ($E_{2}$) to estimate the distribution of each element in $\tilde{y}^{L}$. The decoded $\tilde{y}^{L}$ and $\tilde{y}_{1}^{H}$ will be sent to CRPE ($E_{3}$)  to estimate the distribution of each element in $\tilde{y}^{H}$. Finally, the decoded $\tilde{y}^{H}$ and  $\tilde{y}^{L}$ will be sent to the decoder ($g_{d}$) to obtain the reconstructed image. The inverse GDN is adopted in the Decoder network ($g_{d}$). Note that the latent representations (e.g. $y^{H}$, $y^{L}$) are quantized, which cause some errors during this processing. To address this problem, uniform noises are added to the latent representations in the training stage. The uniform quantization (i.e., round function in this work) is applied to the latent representations during the test. As illustrated in Fig. \ref{framework}, the quantized HR and LR latent representations are entropy-coded using two separate arithmetic encoder and decoder.

In the ablation experiments in Sec. III, we will explore the performance gains of the additional hyper layer and the CRPE modules in our scheme. Since the additional layer does not adopt the context model, the decoding complexity will not increase too much compared to the two-hyper hyper-prior model.

\subsection{Network Architecture}

\begin{figure*}
\centering
\subfigure
{\begin{minipage}[t]{0.5\linewidth}
\centering
\includegraphics[width=\columnwidth]{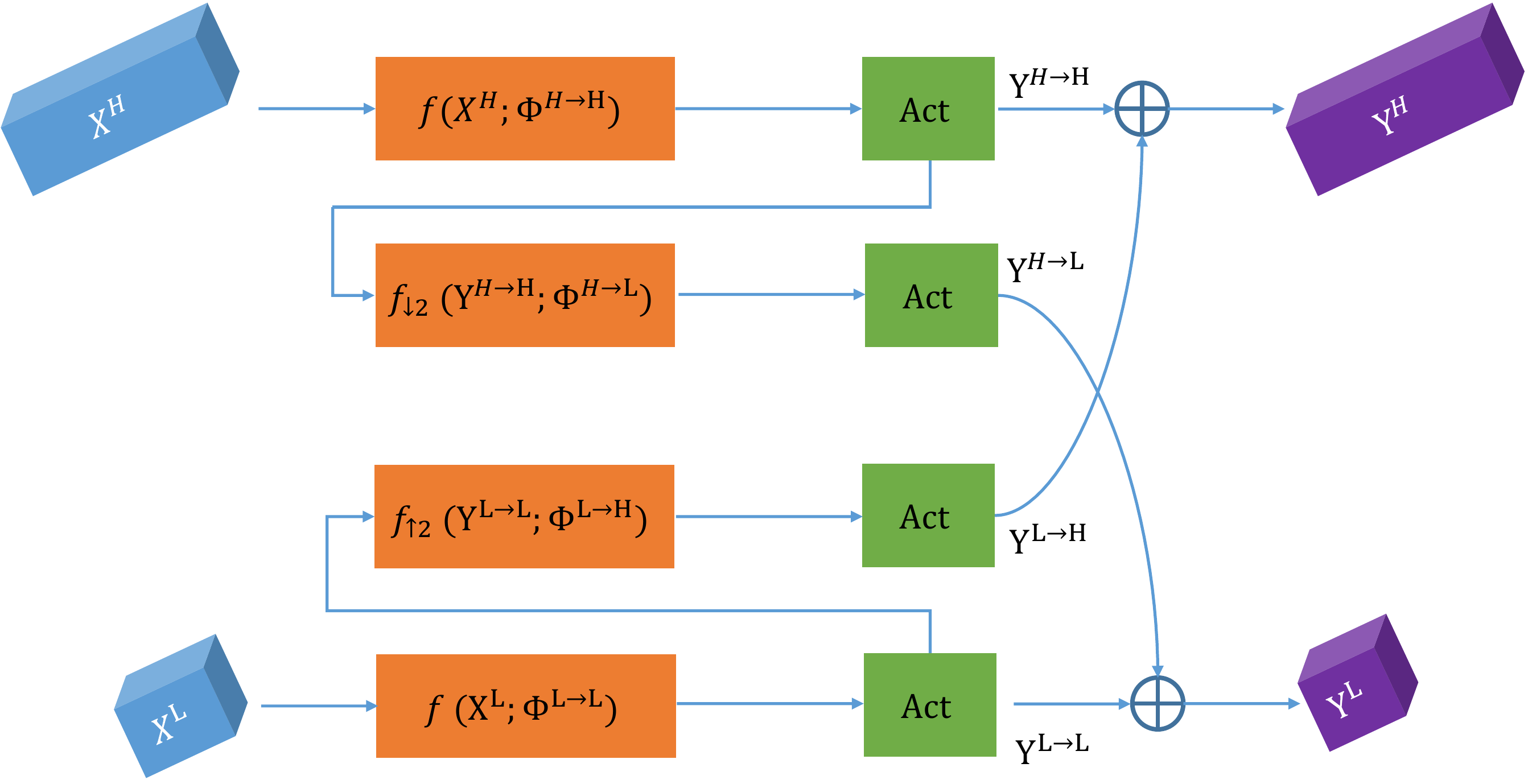}
\end{minipage}
}%
\subfigure
{\begin{minipage}[t]{0.5\linewidth}
\centering
\includegraphics[width=\columnwidth]{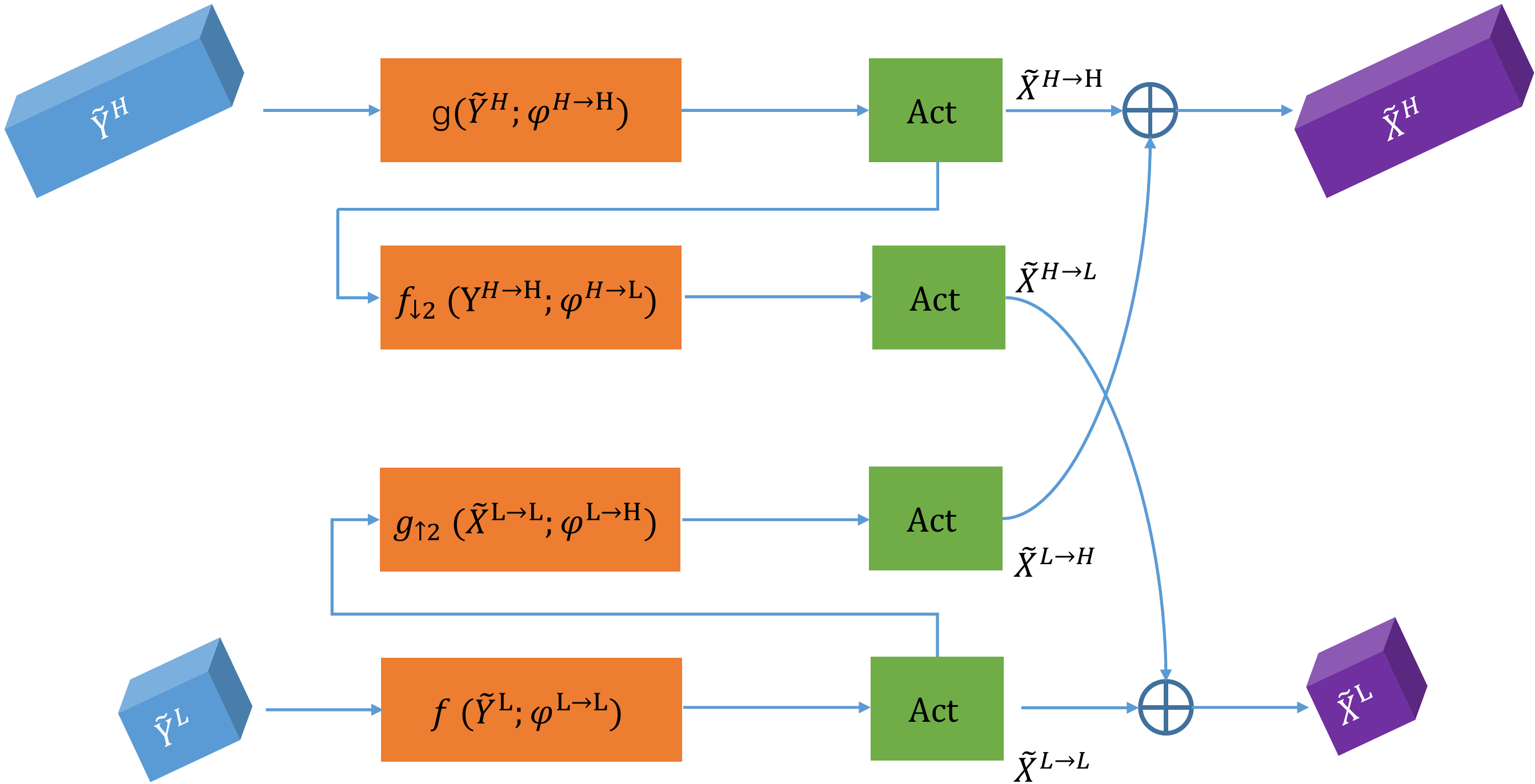}
\end{minipage}
}%
\centering
\caption{Architecture of the generalized octave convolution. $\bigoplus :$ element-wise
addition; Act: the activation layer; $f$: regular vanilla convolution; $g$: regular transposed-convolution; $f_{↓2}$ : regular convolution with stride 2; $g_{↑2}$: regular transposed-convolution with stride 2.}
\label{GoConv}
\end{figure*}

Octave convolution is similar to the wavelet transform \cite{JEPE2000_book}, since it can factorize each channel into HR and LR components.  The generalized octave convolution (GoConv) and generalized octave transposed-convolution (GoTConv) in \cite{Mahammand_AAAI, Lin_MMSP} are shown in Fig. \ref{GoConv}. The strided convolutions are introduced in \cite{Lin_MMSP} to replace the sub-sampling operation in the original octave convolution in \cite{octave}.

The encoder and decoder network structures of the proposed scheme are described in Fig. \ref{framework}. The size of the input image $X$ is denoted as $W\times H\times 3$, where $W$ and $H$  represent the width and height of the image, respectively. In order to make the network easier to train and converge, the input image is normalized to $[-1, 1]$ by calculating($\frac{X}{127.5} - 1.0$).  The analysis transforms and synthesis transforms in the autoencoder layers and the two hyper layers have symmetric networks, apart from using convolutional and de-convolutional filters, respectively.

The GoConv and GoTConv architectures are introduced to the Encoder ($g_{e}$) and Decoder ($g_{d}$). GDN/IGDN transforms are respectively adopted in GoConv and GoTConv  in the first-layer encoder and decoder, while leaky ReLU is utilized in the second and third hyper autoencoder and the CRPE modules.

The encoder ($g_{e}$) of the first layer consists of four GoConv modules, each is followed by down-sampling with a factor of 2. The hyper encoder ($h_{e}$) of the second layer consists of three GoConv modules. The down-sampling factors are set to 1, 2, 2, separately.  The hyper encoder ($f_{e}$) of the third layer is comprised of three GoConv modules, each is followed by a convolution with a stride of 1. The synthesis transform mirrors the analysis transform of the three layers. The last convolution layer with three filters is expected to generate the decompressed image in RGB space.

\subsection{The Cross-Resolution Parameter Estimation (CRPE)}

The CRPE modules including $E_{1}$, $E_{2}$, $E_{3}$ are introduced in our scheme to improve the performance, which contain two types, shown in Fig. \ref{scheme_structure} and Fig. \ref{framework} . The first type is incorporated in high and low resolutions modules. Our first CRPE modules aim to improve the image compression. The second type of the CRPE modules is in different layers, which can combine the LR parts of different layers.  The second type of the CRPE modules are easier to obtain the global information of different layers, which can provide more accurate information for encoding and decoding. These two types of CRPE modules can improve the flow of information, in which the decoded information is provided as side information to the to-be-decoded elements.

Each CRPE consists of several up-sampling convolution layers with a stride of 2 and several normal convolution modules with a stride of 1, where the up-sampling multipliers are determined by the ratio of the width or height of the input two tensors. For example, if the width and height of one input tensor is four times the width and height of another tensor, two up-sampling layers with a stride of 2 are required.

The R-D performance of the learned image compression methods without using the context model will drop significantly. The element being decoded can make full use of the previously decoded neighbor representations. The signal flow of most image compression frameworks is unidirectional. The proposed CRPE can realize the fusion of different information. The  CRPE modules can provide decoded information as additional information to undecoded elements, so that element decoding can utilize more global information instead of local information.

We mainly consider the following two aspects in the process of designing the CRPE to further improve the performance of image compression. First, we decode the LR latent representation instead of decoding low bitrate and high bitrate elements simultaneously in parallel. The LR latent representation is first decoded, and then the decoded LR information is provided to the HR information for decoding. Second, the first hyper decoder can exploit the decoded information of the second hyper layer to decode.

\begin{table*}
\caption{The average Encoding time (s), Decoding time(s) and model size (MB) on Kodak and Tecnick datasets.}
\begin{center}
\begin{tabular}{|c|c|c|c|c|c|}
\hline
\textbf{Dataset}& \textbf{Scheme} & \textbf{Bit Rate} & \textbf{Encoding Time (s)} & \textbf{Decoding Time (s)} & \textbf{Model size (MB)}\\ \hline
\multirow{10}{*}{Kodak}

                         & VVC \cite{VVC} & Low  & 402.27s & 0.60s  & None\\
                         & Lee2019 \cite{Lee_2020} & Low  & 10.721s & 37.88s & 123.8MB \\
                         & Hu2021 \cite{Hu_2021} & Low  &35.72s & 77.33s  & 84.6MB\\
                         & Cheng2020 \cite{cheng2020} &Low &20.89s  &22.14s & 50.8MB \\
                         & Chen2021 \cite{chen2021} &Low &402.26s  &2405.14s & 300.99MB \\
                         & Akbari2021 \cite{Mahammand_AAAI} &Low &73.46s  &75.14s & 74.67MB \\
                         & \textbf{Ours}   & Low   & 6.87s                    & 5.90s   & 77.7MB \\ \cline{2-6}

                         & VVC\cite{VVC}      & High  &760.81s &0.81s   & None \\
                         & Lee2019 \cite{Lee_2020} & High  &21.74s  &70.44s  & 292.6MB\\                     
                         & Hu2021 \cite{Hu_2021}  & High  &70.75s  &207.77s & 290.9MB \\
                         & Cheng2020  \cite{cheng2020} &High  &90.91s  &93.28s  & 175.18MB \\
                        & Chen2021 \cite{chen2021} &High &360.46s  &10653.14s & 300.99MB \\
                         & Akbari2021 \cite{Mahammand_AAAI} &High &186.38s  &190.36s & 218.47MB \\
                         & \textbf{Ours} & High   &7.21s & 6.11s & 282.0MB \\ \hline
\multirow{10}{*}{Tecnick}
                         & VVC \cite{VVC}      & Low  &235.46s &0.857s    & None \\
                         & Lee2019 \cite{Lee_2020}  & Low  &54.8    &138.81    & 123.8MB\\
                         & Hu2021 \cite{Hu_2021}   & Low  &132.93s &661.87s   & 84.6MB \\
                         & Cheng2020 \cite{cheng2020} &Low   &50.49s  &53.24s    & 50.8MB\\
                          & Chen2021 \cite{chen2021} &Low & 1579.60s  &4798.5s & 300.99MB \\
                        & Akbari2021 \cite{Mahammand_AAAI} &Low &155.32s  &159.26s & 74.67MB \\
                         & \textbf{Ours} & Low & 7.33s & 6.29s  & 77.7MB                \\ \cline{2-6}
                         & VVC \cite{VVC}     & High  &2156.59s &1.794s   & None\\
                         & Lee2019 \cite{Lee_2020} & High  &110.42s  &291.60s  & 292.6MB\\
                         & Hu2021 \cite{Hu_2021}  & High  &272.74s  &600.33s  & 290.9MB \\
                         & Cheng2020 \cite{cheng2020} &High  &400.63s  &404.12s  & 175.18MB\\                         & Chen2021 \cite{chen2021} &High &1765.32s  &4983.6s & 300.99MB \\
                        & Akbari2021 \cite{Mahammand_AAAI} &High &645.46s  &648.14s & 218.47MB \\
                         & \textbf{Ours}& High &7.56s & 7.60s   & 282.0MB                 \\ \hline
\end{tabular}
\end{center}
\label{runing_time}
\end{table*}

\subsection{Loss Function}

The objective function for training is composed of two terms: the rate $R$ , which is the expected length of the bitstream, and distortion $D$, which is the expected error between the input and
decompressed images. The tradeoff between the rate and distortion is determined by a Lagrange multiplier denoted by $\lambda$. The R-D optimization problem is then defined as follows:

\begin{equation}\label{vanilla}
\begin{split}
   L = R &+ \lambda D \\
   R &= R^{H}+R^{L}\\
   D &= E_{x \sim P_{x}}[d(x,\hat{x})]
\end{split}
\end{equation}

where $p_{x}$ represents the unknown distribution of the natural images. $D$ indicates the  distortion metric such as the mean squared error (MSE) or MS-SSIM. $R^{H}$ and $R^{l}$ are the rates corresponding to the HR and LR parts of the output bitstream.

To increase the LF part in the whole bit stream during the training of the multi-layer network, an additional information-fidelity loss is presented. If the information-fidelity loss function is not adopted in the proposed framework, the LR part will only take a very small portion in the final bit stream, which does not take full advantage of the multi-resolution feature. This problem can be effectively addressed by introducing the information-fidelity loss function. This loss term encourages the LR module to maintain the LR critical information during training, which is formulated as:

\begin{equation}\label{vanilla}
\begin{split}
L_{if}  = \lambda_{1} {\parallel F(y^{L},\theta) - y^{L} \parallel}^2 + \lambda_{2} {\parallel F({y_{1}}^{L},\omega) - {y_{1}}^{L} \parallel}^2
\end{split}
\end{equation}

The function $F$ with trainable parameter $\theta$ and $\omega$ is one convolutional layer with no non-linear activation.  By including $y^{L}$ and ${y_{1}}^{L}$ in the cost function, more LR bits can be produced by the optimized networks. The parameters $\lambda_{1}$ and $\lambda_{2}$ are scaling factors. Ablation experiments will verify the role of the information-fidelity loss. Therefore, the final loss function is as follows:
\begin{equation}\label{vanilla}
\begin{split}
L  = R + \lambda D + L_{if}
\end{split}
\end{equation}

\begin{figure*}
\centering
\subfigure
{\begin{minipage}[t]{0.5\linewidth}
\centering
\includegraphics[width=\columnwidth]{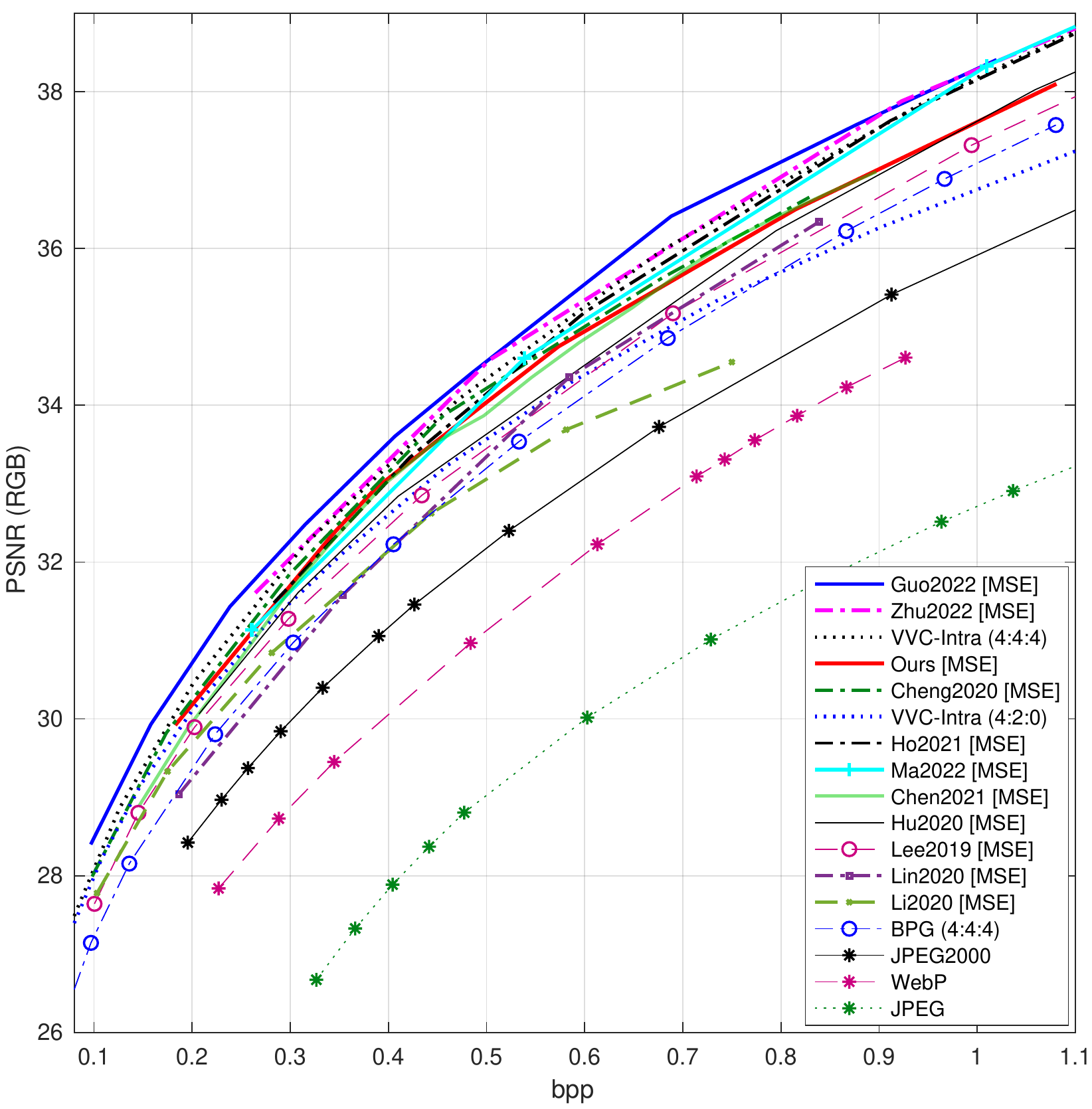}
\end{minipage}
}%
\subfigure
{\begin{minipage}[t]{0.5\linewidth}
\centering
\includegraphics[width=\columnwidth]{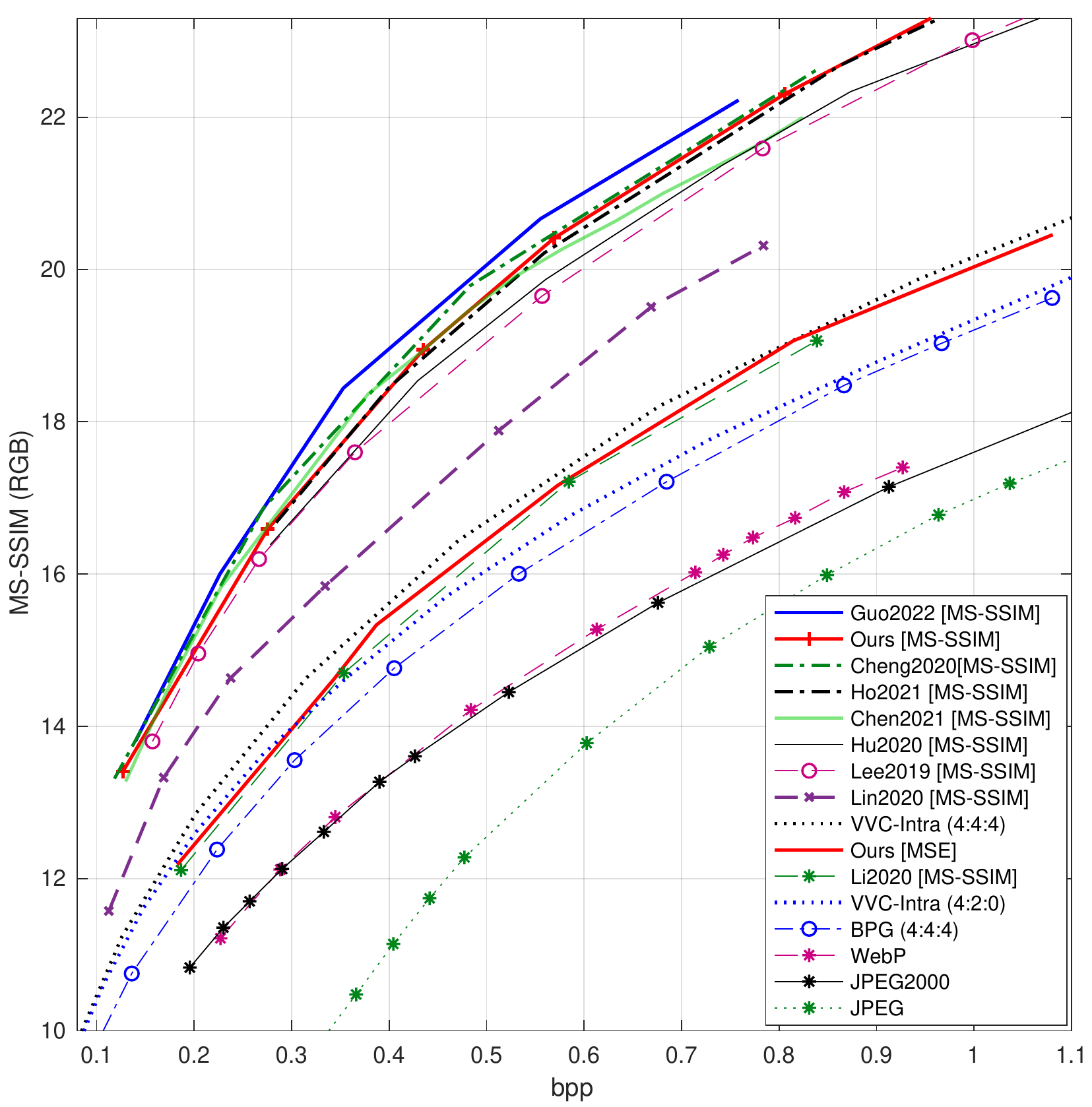}
\end{minipage}
}%
\centering
\caption{Average  comparison results on all 24 Kodak images on both PSNR and MS-SSIM.}
\label{test_kodak}
\end{figure*}

\begin{figure*}
\centering
\subfigure
{\begin{minipage}[t]{0.5\linewidth}
\centering
\includegraphics[width=\columnwidth]{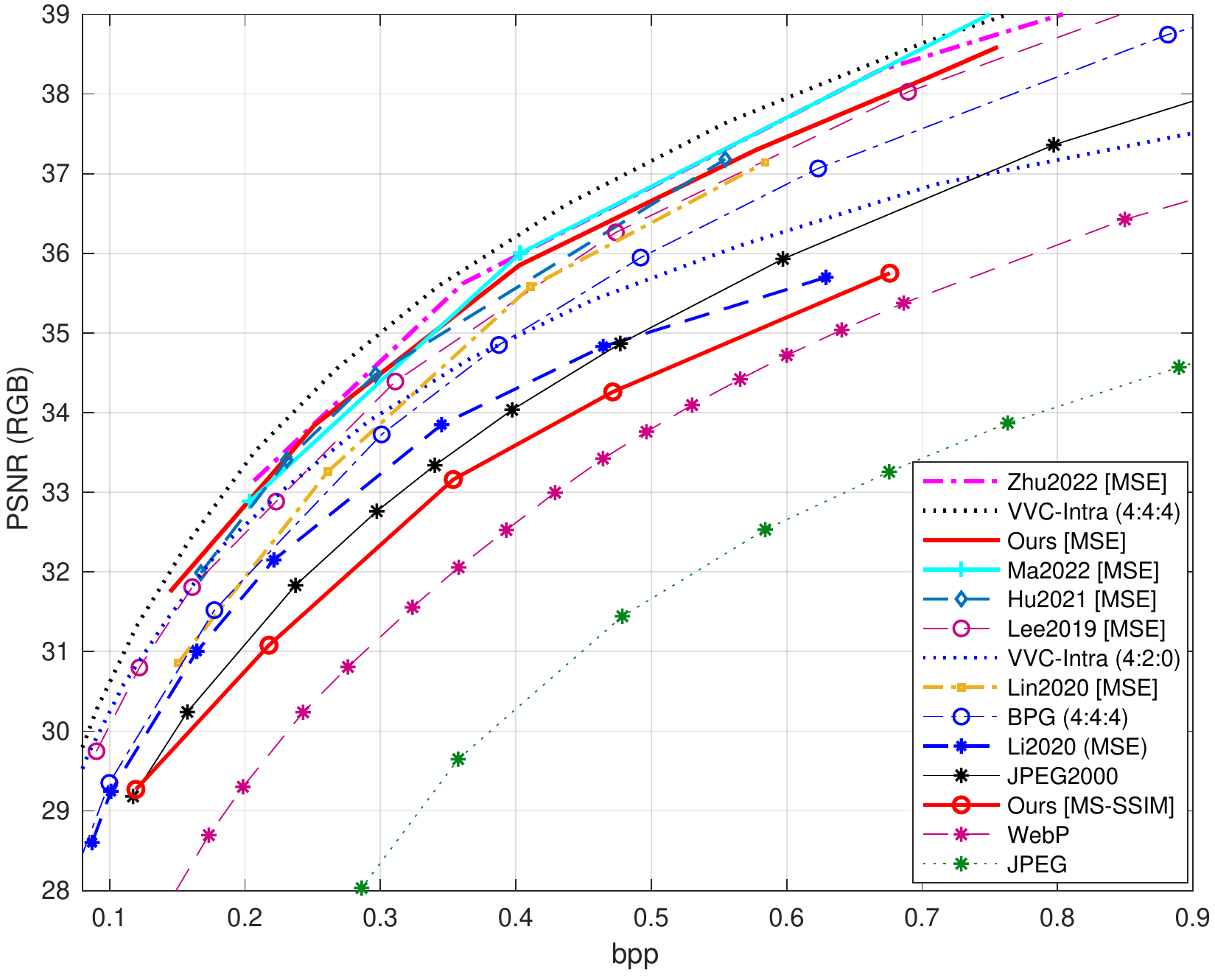}
\end{minipage}
}%
\subfigure
{\begin{minipage}[t]{0.5\linewidth}
\centering
\includegraphics[width=\columnwidth]{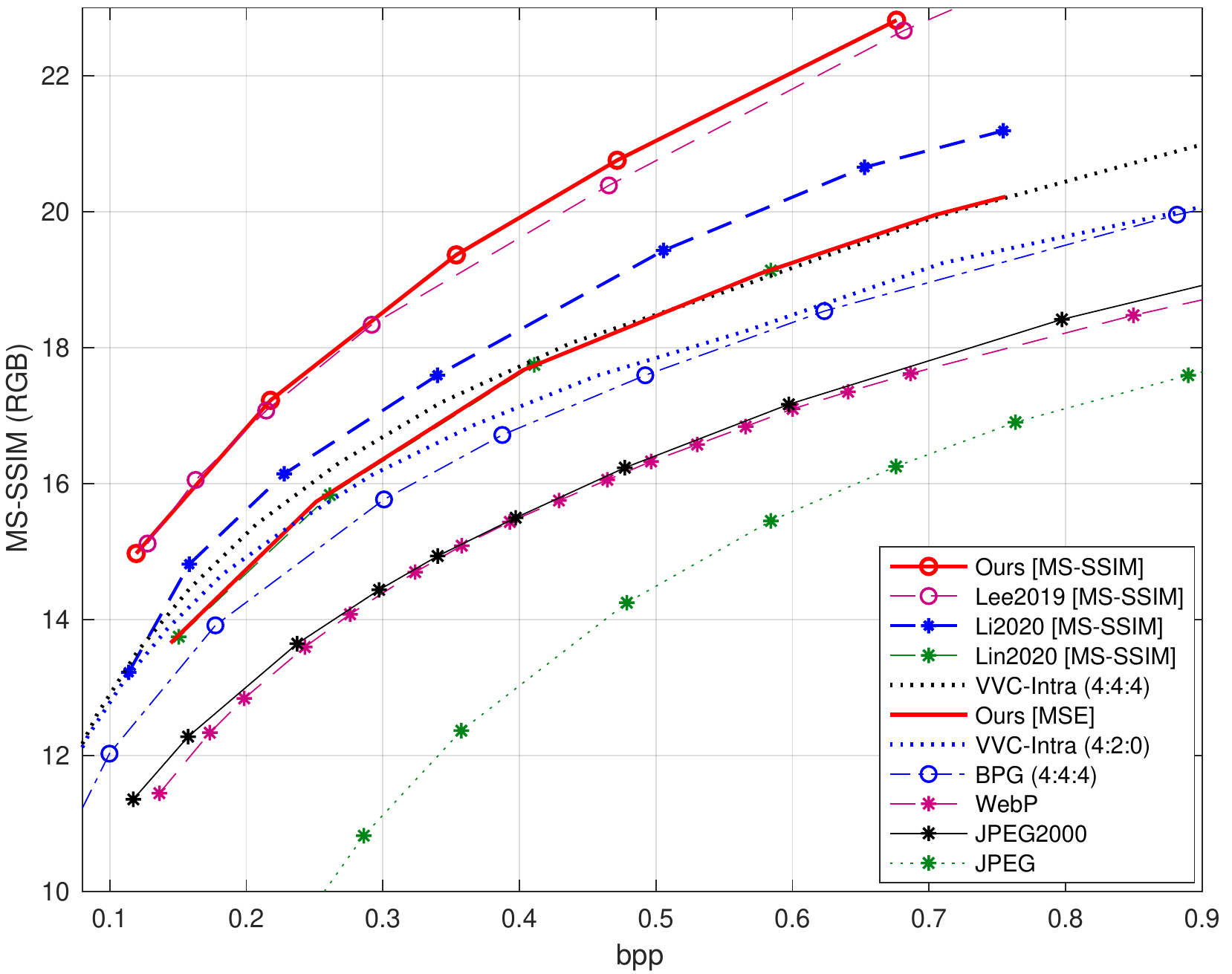}
\end{minipage}
}%
\centering
\caption{Average evaluation results on all 100 Tecnick images on both PSNR and MS-SSIM.}
\label{test_Tecnick}
\end{figure*}

\section{Experiments}
\label{Experiment}

\begin{figure*}
\centering
\subfigure[Original]{
\begin{minipage}[t]{0.33\linewidth}
\centering
\includegraphics[scale=0.65]{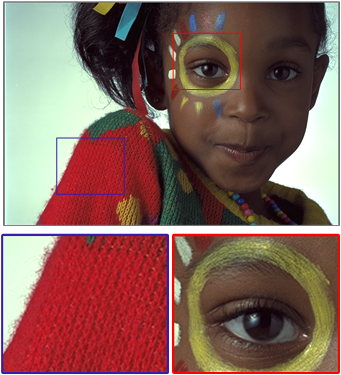}
\end{minipage}
}%
\subfigure[JPEG(0.166/21.82/0.737)]{
\begin{minipage}[t]{0.33\linewidth}
\centering
\includegraphics[scale=0.65]{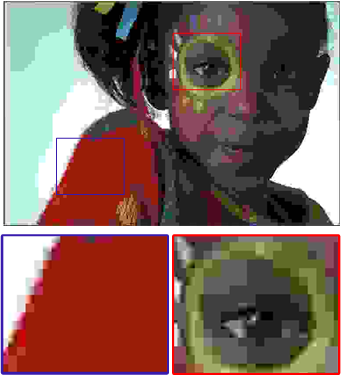}
\end{minipage}
}%
\subfigure[JPEG2000(0.116/29.14/0.924)]{
\begin{minipage}[t]{0.33\linewidth}
\centering
\includegraphics[scale=0.65]{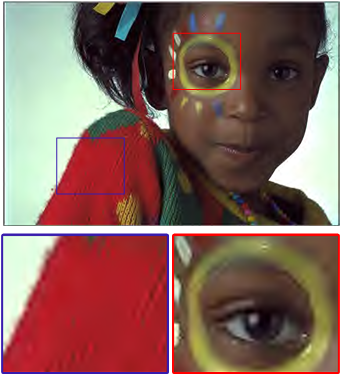}
\end{minipage}
}%

\subfigure[Webp(0.112/28.460/0.919)]{
\begin{minipage}[t]{0.33\linewidth}
\centering
\includegraphics[scale=0.65]{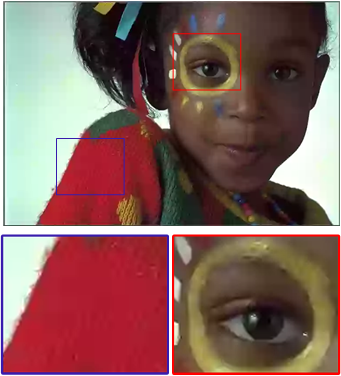}
\end{minipage}
}%
\subfigure[BPG(4:4:4)(0.101/30.33/0.934)]{
\begin{minipage}[t]{0.33\linewidth}
\centering
\includegraphics[scale=0.656]{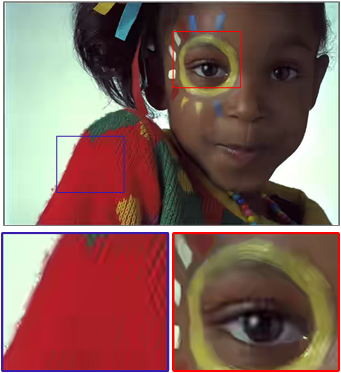}
\end{minipage}
}%
\subfigure[Ours(0.101/30.60/0.944)]{
\begin{minipage}[t]{0.33\linewidth}
\centering
\includegraphics[scale=0.65]{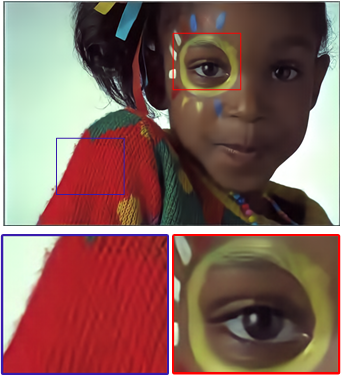}
\end{minipage}
}%
\centering
\caption{Example 1 in the Kodak dataset (bpp, PSNR(dB), MS-SSIM).}
\label{Example1}
\end{figure*}

\begin{figure*}
\centering
\subfigure[Original]{
\begin{minipage}[t]{0.33\linewidth}
\centering
\includegraphics[scale=0.65]{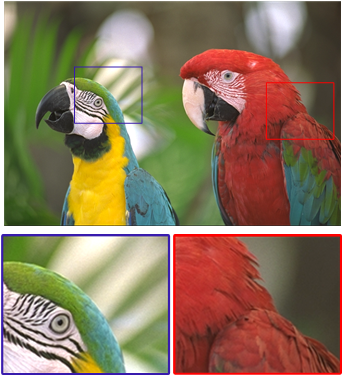}
\end{minipage}
}%
\subfigure[JPEG(0.159/22.53/0.726)]{
\begin{minipage}[t]{0.33\linewidth}
\centering
\includegraphics[scale=0.65]{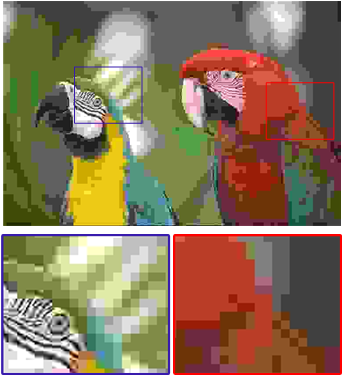}
\end{minipage}
}%
\subfigure[JPEG2000(0.0997/30.73/0.933)]{
\begin{minipage}[t]{0.33\linewidth}
\centering
\includegraphics[scale=0.65]{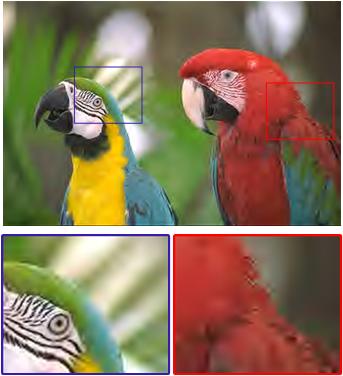}
\end{minipage}
}%

\subfigure[Webp(0.098/28.94/0.920)]{
\begin{minipage}[t]{0.33\linewidth}
\centering
\includegraphics[scale=0.65]{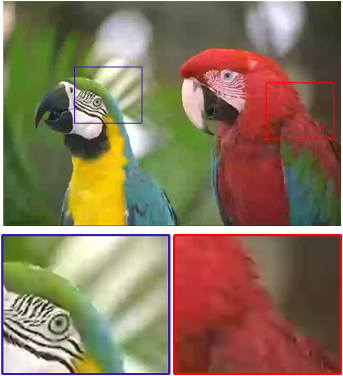}
\end{minipage}
}%
\subfigure[BPG(4:4:4)(0.0999/32.29/0.950)]{
\begin{minipage}[t]{0.33\linewidth}
\centering
\includegraphics[scale=0.65]{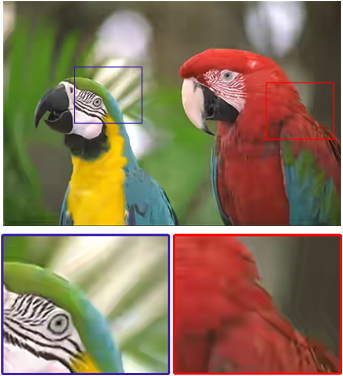}
\end{minipage}
}%
\subfigure[Ours(0.0945/32.63/0.956)]{
\begin{minipage}[t]{0.33\linewidth}
\centering
\includegraphics[scale=0.65]{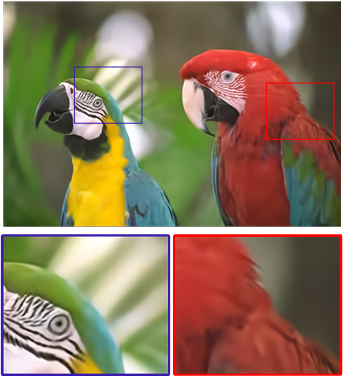}
\end{minipage}
}%
\centering
\caption{Example 2 in the Kodak dataset (bpp, PSNR(dB), MS-SSIM).}
\label{Example2}
\end{figure*}

\begin{figure*}
\centering
\subfigure[Original]{
\begin{minipage}[t]{0.33\linewidth}
\centering
\includegraphics[scale=0.625]{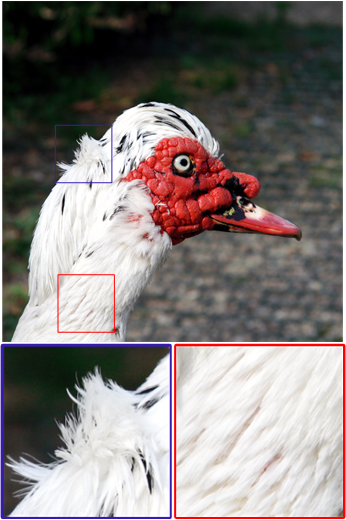}
\end{minipage}
}%
\subfigure[JPEG(0.142/25.353/0.767)]{
\begin{minipage}[t]{0.33\linewidth}
\centering
\includegraphics[scale=0.625]{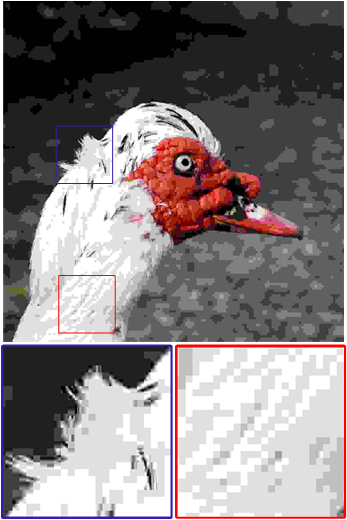}
\end{minipage}
}%
\subfigure[JPEG2000(0.066/32.939/0.953)]{
\begin{minipage}[t]{0.33\linewidth}
\centering
\includegraphics[scale=0.625]{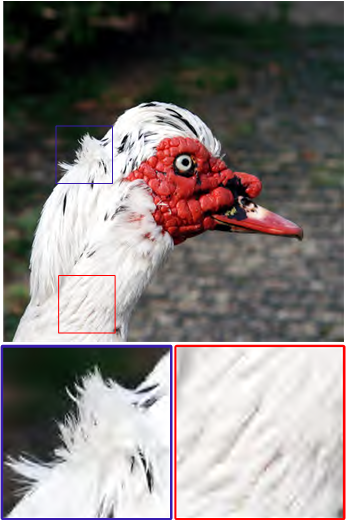}
\end{minipage}
}%

\subfigure[Webp(0.072/30.528/0.0.934)]{
\begin{minipage}[t]{0.33\linewidth}
\centering
\includegraphics[scale=0.625]{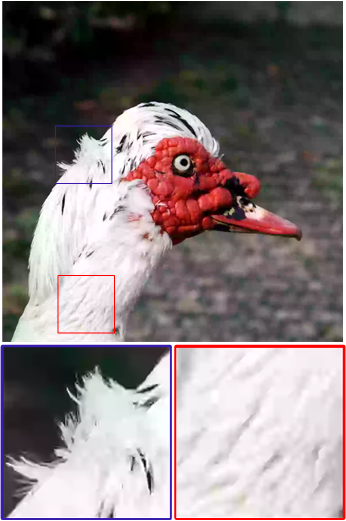}
\end{minipage}
}%
\subfigure[BPG(4:4:4)(0.062/33.469/0.953)]{
\begin{minipage}[t]{0.33\linewidth}
\centering
\includegraphics[scale=0.625]{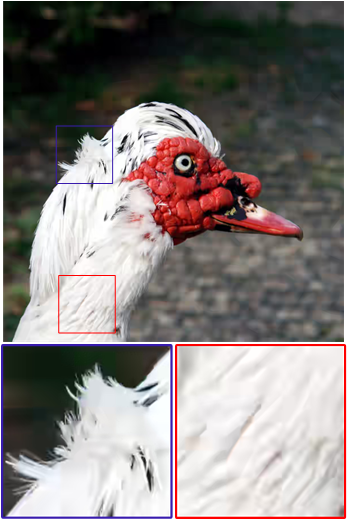}
\end{minipage}
}%
\subfigure[Ours(0.064/34.18/0.961)]{
\begin{minipage}[t]{0.33\linewidth}
\centering
\includegraphics[scale=0.625]{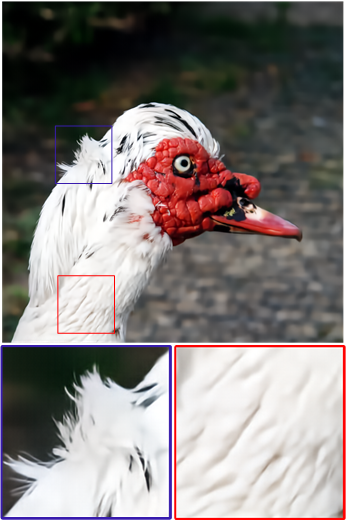}
\end{minipage}
}%
\centering
\caption{Example 3 in the Tecnick dataset (bpp, PSNR(dB), MS-SSIM).}
\label{Example3}
\end{figure*}

In this section, we evaluate the performance of the proposed method on the Kodak PhotoCD dataset \cite{Kodak} and Tecnick dataset \cite{Tecnick}. The Kodak PhotoCD dataset is composed of 24 lossless PNG format images. And the resolution of each test image mainly contains two types: $768 \times 512$ and $512 \times 768$. The Tecnick dataset contains 100 lossless PNG images, in which each test image have a resolution of $1200 \times 1200$. We compare with some recent learning compression methods, including Zhu2022 \cite{Zhu2022}, Ma2022 \cite{Ma_2022_PAMI},  Guo2022\cite{Guo_2022}, Ho2021 \cite{Ho2021}, Chen2021 \cite{chen2021}, Cheng2020 \cite{cheng2020}, Lee2019 \cite{Lee_2020}, Lin2020 \cite{Lin_MMSP}, Hu2021 \cite{Hu_2021}, Li2020 \cite{Content_Weighted}, VVC (4:4:4) \cite{VVC}, VVC (4:2:0) \cite{VVC},  BPG (4:4:4) \cite{BPG}, JPEG2000, WebP, JPEG. The origin MS-SSIM values are mapped to decibels $-10 \log_{10}(1 - MS\textnormal{-}SSIM) $ to better distinguish the performance difference.

\subsection{Training Set}
 The CLIC dataset \cite{CLIC} and LIU4K dataset \cite{LIU_dataset} are used to train our network. All the training images are resized to $2000 \times 2000$ to train our models.  Data augmentation approaches including rotation and scaling were introduced to randomly extract 81,650 patches with $384 \times 384$. The patches are saved as a lossless PNG format.

\subsection{Parameter Settings}

We optimized the proposed models using two evaluation metrics including mean square error (MSE) and MS-SSIM \cite{MS-SSIM}. When optimized for MSE metric, the parameter $\lambda$ is selected from the set $\{0.003,0.008,0.02,0.04,0.07\}$, each trains a model for a different bit rate. The number of filters $N$ is set to 256 for two lower bit rates, and is set to 448 for the three higher bit rates. When the MS-SSIM metric is used, the parameter $\lambda$ is in the set $\{0.003,0.008, 0.0015, 0.02, 0.04\}$.  The number of filters $N$ is set to 256 for two lower bit rates, and is set to 448 for the three higher bit rates.  In the GoConv part, half of the latent representations have low resolutions, and the other half have high resolutions. 

The network training process can be divided into three stages. In the first stage, we pretrain the main encoder $g_{e}$ and decoder $g_{d}$ to obtain a good reconstructed images. The parameters of hyper encoders and decoders are not involved in training. In the second stage, we randomly initialize the hyper encoder $f_{e}$ and the decoder $f_{d}$. The parameters of the $f_{e}$ and $f_{d}$ added to training parameters to jointly train the model. In third stage, we randomly initialize the hyper encoder $h_{e}$ and the decoder $h_{d}$, the whole model is trained in an end-to-end manner.  Each model is  trained up to $150,000$ iterations for each $\lambda$ to obtain stable performance. The Adam solver with a batch size of 8 is adopted and the learning rate is set to 0.0004 during the training.

\subsection{Encoding and Decoding Complexity}

Table \ref{runing_time} compares the complexities of different methods. Since VVC, Hu2020 \cite{Hu_AAAI}, and cheng2020 \cite{cheng2020} only run on CPU, to ensure fairness, we evaluate the running time of different methods on an 2.9GHz Intel Xeon Gold 6226R CPU. The average time at low bit rate and high bit rate over all images on Kodak dataset is used. The average model sizes at low bit rates and high bit rates are also reported.

It can be seen from Table \ref{runing_time} that compared to some learned methods, the proposed scheme achieves better results while providing nearly 4 times and 15 times improvement in decoded runtime at low bit rates and high bit rates, respectively (i.e., a decoding time savings of 73.35 $\%$ and 93.44 $\%$ at low bit rates and high bit rates compared with that in Cheng2020 \cite{cheng2020}) on Kodak dataset, respectively.  Our model size is slightly bigger than Cheng2020 \cite{cheng2020} and Akbari2021 \cite{Mahammand_AAAI}, and also are smaller than other learned methods. Compared to Chen2021 \cite{chen2021},  the proposed scheme  provides nearly 50 times and 400 times improvement in encoding and decoded runtime at low bit rates and high bit rates on Kodak dataset, respectively. Compared to  Akbari2021 \cite{Mahammand_AAAI},  the proposed scheme  provides nearly 10 times and 25 times improvement in encoding and decoded runtime at low bit rates and high bit rates on Kodak dataset, respectively. Compared to VVC, our encoding is about 58 times and  105 times faster on Kodak dataset, but the decoding is much slower.  Our scheme has greater advantages in encoding time and decoding time in Tecnick dataset compared with other learned methods. Therefore the proposed scheme achieves the new state of the art in learned image coding when considering both the time complexity and space complexity.

\subsection{Comparison}

The average MS-SSIM and PSNR performance over the 24 Kodak images are illustrated in Fig. \ref{test_kodak}. Guo2022 \cite{Guo_2022} achieves best results and outperforms VVC (4:4:4) in PSNR. Zhu2022 \cite{Zhu2022} is sightly worse than Guo2022 \cite{Guo_2022} and it still obtains better performance than VVC (4:4:4) in PSNR. Ho2021 \cite{Ho2021}, Ma2022 \cite{Ma_2022_PAMI}, and our method achieve the same performance at low bit rates. Ma2022 \cite{Ma_2022_PAMI} is sightly better than  Ho2021 \cite{Ho2021} and our proposed method. The proposed scheme outperforms Lee2019 \cite{Lee_2020}, Lin2020 \cite{Lin_MMSP}, Chen2021 \cite{chen2021}, Hu2021 \cite{Hu_AAAI}, Li2020 \cite{Content_Weighted}, VVC (4:2:0) \cite{VVC}, BPG (4:4:4), JPEG2000, WebP, JPEG in terms of both PSNR and MS-SSIM metrics. When optimized for PSNR, VVC (4:4:4) achieves the best result. Cheng2020 \cite{cheng2020} achieves best result in learned methods.  Our proposed method almost achieves the same performance with Cheng2020 \cite{cheng2020}, and it is slightly worse than VVC (4:4:4) \cite{VVC} at low bit rates.  When optimized for MS-SSIM, Gu2020 \cite{Guo_2022} optimized for MS-SSIM achieves the best performance in all methods. Cheng2020 \cite{cheng2020} is sightly worse than Gu2020 \cite{Guo_2022}. The proposed MS-SSIM method and Ho2021 \cite{Ho2021} almost obtain the same performance with Cheng2020 \cite{cheng2020} at high bit rates and achieve slightly worse performance than \cite{cheng2020} at low bit rates. We calculate the BD-rate ($\%$) of different methods including Chen2021 \cite{chen2021}, Cheng2020 \cite{cheng2020}, Lee2019 \cite{Lee_2020},  Hu2021 \cite{Hu_2021}, Akbari2021 \cite{Mahammand_AAAI} and GLLMM \cite{GLLMM}. The best traditional codec VVC (4:4:4) \cite{VVC} is considered as anchor.  Since some methods are not available, we do not compare them. The result is illustrated in Table \ref{BD_rate}.  From Table \ref{BD_rate}, we can find that the GLLMM achieves an average of 4.7 $\%$ bitrate saving over VVC (4:4:4) on Kodak dataset. Our method is very close to  VVC (4:4:4) and has 5.6 $\%$ bitrate increase.

\begin{table*}
\caption{The BD-rate ($\%$) of different methods. The
negative numbers represent bitrate saving and positive numbers
indicate bitrate increase.}
\begin{center}
  \begin{tabular}{cccccccc}
  \hline
  \textbf{Methods}  &  \textbf{Chen2021 \cite{chen2021}}& \textbf{ Cheng2020 \cite{cheng2020}}& \textbf{Lee2019 \cite{Lee_2020}}& \textbf{Hu2021 \cite{Hu_2021}} & \textbf{Akbari2021 \cite{Mahammand_AAAI}} & \textbf{GLLMM\cite{GLLMM}} & \textbf{Ours} \\
  \hline
  \textbf{BD-rate} &8.6 &4.8 &17.0 &11.1 &9.4 &-3.13 &5.6\\
  \hline
\end{tabular}
\label{LR_HR}
\end{center}
\label{BD_rate}
\end{table*}

We also evaluate the performance of the proposed method on the Tecnick dataset compared to traditional methods and some latest learned methods in Fig. \ref{test_Tecnick}. Some methods are not included because we do not have their results. When optimized for PSNR, Zhu2022 \cite{Zhu2022} achieves best results at low bit rates and it is worse than Ma2022 \cite{Ma_2022_PAMI} at high bit rates. Ma2022 \cite{Ma_2022_PAMI} is sightly worse than our method at low bit rates and it achieves better performance than our method at high bit rates. Our method outperforms Lee2019 \cite{Lee_2020}, Li2020PAMI\cite{Content_Weighted}, Lin2020 \cite{Lin_MMSP}, VVC (4:2:0),  BPG(4:4:4), JPEG2000, WebP, JPEG  in term of both PSNR  at all rate points.   When optimized for MS-SSIM, our method achieves the best results than traditional methods and learned methods across a wide bit rates.

To demonstrate the proposed method can generate more visually pleasant results, three visual examples from the Kodak and Tecnick dataset are illustrated in Fig. \ref{Example1}, Fig. \ref{Example2}, and Fig. \ref{Example3}, in which our proposed results are compared with JEPG, JPEG2000 and webp, BPG (4:4:4 chroma format). As seen in the two examples, our proposed method achieves the highest visual quality compared to other codecs, in which more detailed information can be generated. Due to the blocking artifacts, the JPEG has the worst visual performance. The JPEG2000 achieves poor performance owing to the ringing artifacts. The WebP result is blurred in certain regions. The BPG codec achieves a quite smooth result, but the details and fine structures are not preserved in some regions.


\begin{figure}
	\flushleft
		\includegraphics[scale=0.65]{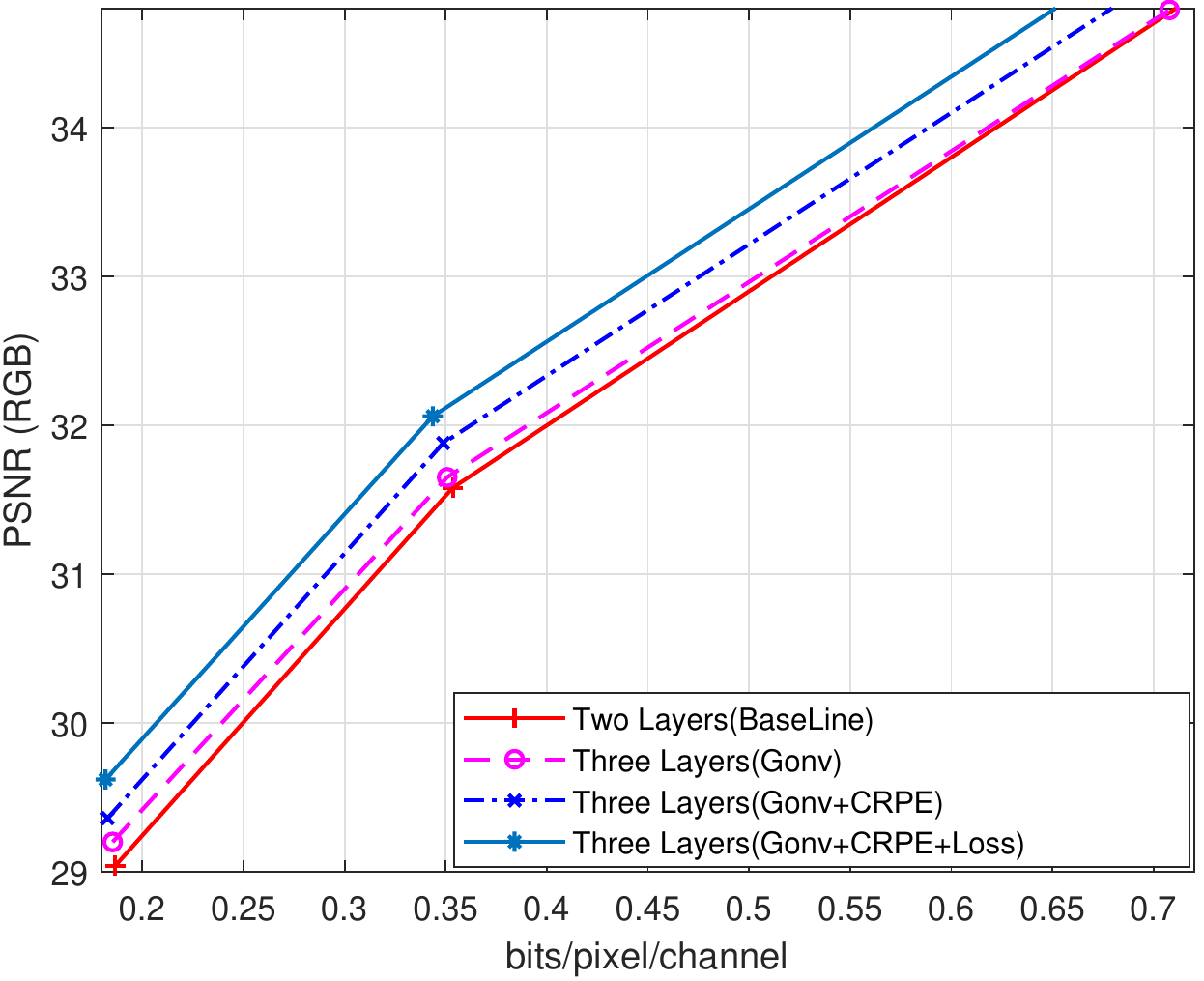}
	\caption{The R-D  curves of several variants of the proposed modules on the Kodak dataset. \textbf{GoConv} represents the  modulated generalized octave convolution
module. \textbf{CPRE} indicates the cross-resolution parameter estimation module. \textbf{Loss} denotes the information-fidelity loss.}
	\label{Ablation}
\end{figure}

\begin{table}
\caption{The proportion of Low bit rates and  the high bit rates.}
\begin{center}
  \begin{tabular}{cccccc}
  \hline
  \textbf{Name}  &  \textbf{Total BPP}& \textbf{Low BPP}& \textbf{High Bpp}& \textbf{Ratio}\\
  \hline
  \textbf{scheme 1} &0.1790 &0.0169 &0.1620&1:9.5858\\
  \textbf{scheme 2} &0.1785 &0.0908 &0.0877&1:0.9658\\
  \textbf{scheme 1} &0.3440 &0.0360 &0.3080&1:8.556\\
  \textbf{scheme 2} &0.3390 &0.1413 &0.1977&1:1.399\\
  \textbf{scheme 1} &0.6850 &0.0991 &0.5859&1:5.914\\
  \textbf{scheme 2} &0.6817 &0.2073 &0.4744&1:2.288\\
  \hline
\end{tabular}
\label{LR_HR}
\end{center}
\label{running_time}
\end{table}

\subsection{The Different Structure Of The Two Different Hyper Layers}

\begin{table*}
\caption{The different hyper structures of different schemes. HE denotes Hyper Encoder, and HD represent Hyper Decoder.}
\begin{center}
  \begin{tabular}{cccccccc}
  \hline
  \textbf{Scheme 1}& \textbf{First Layer} & \textbf{Second Layer}& \textbf{Third Layer} & \textbf{Scheme 2}& \textbf{First layer} & \textbf{Second layer}& \textbf{Third layer} \\
  \hline
  \textbf{HE ($h_{e}$)} &  $3*3, N, s1 $ & $3*3, N, s2\downarrow $ & $3*3, N, s2\downarrow $ &
  \textbf{HE ($h_{e}$)} & $3*3, N, s1 $ & $3*3, N, s2\downarrow $ & $3*3, N, s1 $  \\
  \textbf{HE ($f_{e}$)} & $3*3, N, s2$ & $3*3, N, s1 $ & $3*3, N, s1 $ &
  \textbf{HE ($f_{e}$)} & $3*3, N, s1 $ & $3*3, N, s2\downarrow $ & $3*3, N, s1 $ \\
  \textbf{HD ($f_{d}$)} & $3*3, N, s1 $ & $3*3, N, s1 $ & $3*3, N, s1 $ &
  \textbf{HD ($f_{d}$)} & $3*3, N, s1 $ & $3*3, N, s2\uparrow $ & $3*3, N, s1 $ \\
  \textbf{HD ($h_{d}$)} & $3*3, N, s1 $ & $3*3, N, s2\uparrow $ & $3*3, N, s2\uparrow $ &
  \textbf{HD ($h_{d}$)} & $3*3, N, s1 $ & $3*3, N, s2\uparrow $ & $3*3, N, s1 $ \\
  \hline
\end{tabular}
\label{hyper_structure}
\end{center}
\end{table*}

\begin{table}
\caption{The different structure of two different hyper layers on Kodak dataset.}
\begin{center}
  \begin{tabular}{cccccc}
  \hline
  \textbf{Name}  &  \textbf{N}& \textbf{$\lambda$}&\textbf{Bpp}& \textbf{PSNR(dB)}& \textbf{MS-SSIM}\\
  \hline
  \textbf{scheme 1} &128 &0.002 &0.1343&28.36&0.9224\\
  \textbf{scheme 2} &128 &0.002 &0.1442&28.12&0.9204\\
  \textbf{scheme 1} &256 &0.002 &0.1311&28.22&0.9224\\
  \textbf{scheme 2} &256 &0.002 &0.1977&28.14&0.9212\\
  \hline
\end{tabular}
\label{different_bit_rate}
\end{center}
\label{different_downsample}
\end{table}

The first ablation study is conducted to prove the effectiveness of different components in our proposed framework in Fig. \ref{Ablation}. In this experimental setting, we investigate the effectiveness of the proposed modules including the additional hyper layer, the cross-resolution parameter estimation (CRPE) modules, and  the information-fidelity loss. The original two-layer hyper-prior model with the modulated generalized octave convolution module is considered as the baseline method.  Based on the baseline, we combined different modules to the baseline to investigate the performance improvement level of different modules. To compare as fair as possible, the parameter $\lambda$ is chosen in the set $\{0.003,0.008,0.03\}$. The number of filters $N$ is set to 128 for all bit rates. The PSNR metric is optimized as the distortion function.

As seen in Fig. \ref{Ablation}, the baseline model achieves poor performance. We first add an additional hyper layer to the baseline. Compared to the baseline, the Three Layers (GoConv) slightly improves the R-D  performance.  Then we add three  CRPE modules to the Three Layers (GoConv+CPRE) scheme.  We can find the CRPE modules significantly improves the compression performance. Last, we add the information-fidelity loss to the Three Layers method with CRPE modules (GoConv+CPRE) scheme. The information-fidelity loss slightly improves image compression performance. However, it significantly increases the LR portion in of the whole bit stream.

We compare the bit rate allocation of these two schemes. The comparison results are shown in Table \ref{LR_HR}.

The Three Layers (GoConv+CPRE) and the Three Layers (GoConv+CPRE+Loss) are denoted as scheme 1 and scheme 2, respectively.  We find if the information-fidelity loss function is adopted in the proposed framework, the ratio of LR bit rates improves a lot.

We also explore the impact of the different hyper structures on the image compression performance of the proposed method.  Scheme 1 is the proposed method that the hyper encoder ($h_{e}$) have two downsampling layers.  Scheme 2 is that the hyper encoder ($h_{e}$) and hyper encoder ($h_{e}$) both have one down-sampling layer. The detailed structures are shown in Table \ref{hyper_structure}. $3\times3$, $N$, $s2(s1)$, $\downarrow(\uparrow)$ indicates  the size of the convolution kernel, the number filters, convolution step size, down-sampling (up-sampling) operation, separately. We train four sets of the models.

The detailed parameters and result are shown in Table \ref{hyper_structure} and Table \ref{different_downsample}. First, scheme 1 outperforms  scheme 2 in terms of both PSNR and MS-SSIM. The hyper encoder ($h_{e}$) just have the one down-sampling layer, the latent representations of the $h_{e}$ will occupy a larger bit rate ratio than  $h_{e}$ which have two down-sampling layers. The hyper layer is just an auxiliary layer, which can learn  side information to capture spatial dependencies among the elements of the latent representations in Encoder ($g_{e}$). Therefore, there is no need to spend too many bits to encode the latent representations of the $h_{e}$. Second, at low bit rates, the image compression performance will not improve significantly by increasing the number of filters.  It only takes a few channels to learn the relevant features of the image. However, as the bit rate increases, if the number of channels is not increased, the image compression performance will decrease significantly. To achieve better R-D performance, most approaches used different channel numbers to train models to obtain different bit rates.

\section{Conclusion}
\label{Conclusion}

As mentioned before, the learned methods based on context-adaptive entropy models achieve state-of-the-art performance among all the learned methods, in which the hyper-prior and autoregressive models are jointly optimized to significantly remove the spatial dependencies of the latent representations. In this paper, we presented a multi-layer hyper-prior and cross-resolution parameter estimation image compression method based on generalized octave convolution. We remove the context models from the image compression to improve the parallelism of decoding, which will reduce the  R-D performance and lead to a strong spatial correlation in latent representations. We improve the performance of the compression network from three aspects. First, the generalized octave convolution was introduced to factorize the latent representations into high and low resolution. Second, an additional hyper-layer is introduced to further remove the spatial redundancy of the latent representation. Third, to allow the undecoded components to utilize the previously decoded information as much as possible, the cross-resolution parameter estimator (CRPE) modules are introduced to improve the R-D curve performance.

Experiments show that the proposed scheme without context models can outperform VVC (4:2:0) and some recent learned methods in terms of both PSNR and MS-SSIM metrics across a wide bit rates. However, the proposed scheme can still has some room to improve. For example, we can exploit more efficient probability models to accurately establish the probability distribution of potential representations. The complexity of the proposed method can be further optimized by different approaches such as model compression and optimization.



\bibliographystyle{cas-model2-names}

\bibliography{references}

\end{document}